# Distance based (DBCP) Cluster Protocol for Heterogeneous Wireless Sensor Network

Surender Kumar
Department of Computer Engg.UPES, Dehradun
INDIA

Manish Prateek
Department of Computer Engg, UPES, Dehradun
INDIA

Bharat Bhushan
Department of Computer App.
Khalsa College, Yamunanagar
INDIA

## ABSTRACT

Clustering is an important concept to reduce the energy consumption and prolonging the life of a wireless sensor network. In heterogeneous wireless sensor network some of the nodes are equipped with more energy than the other nodes.

Many routing algorithms are proposed for heterogeneous wireless sensor network. Stable Election Protocol (SEP) is one of the important protocol in this category. In this research paper a novel energy efficient distance based cluster protocol (DBCP) is proposed for single hop heterogeneous wireless sensor network to increase the life and energy efficiency of a sensor network. DBCP use the average distance of the sensor from the base station as the major issue for the selection of a cluster head in the sensor network.

## General Terms

Wireless Sensor Networks, Routing Protocol, Heterogeneous Network, Clustering,

## 1. INTRODUCTION

Recent technological advances in hardware and communication technology enable the development of tiny sensors with limited signal processing and wireless communication [9]. Sensor network consists of hundreds or thousands of unattended sensors which are randomly deployed in the area of interest. Wireless sensor network have made reach in many critical applications like battlefield surveillance, habitat monitoring, transportation traffic monitoring [1], [2], [3].

Sensor nodes are highly energy constrained devices as they are battery operated devices and due to harsh deployment of nodes it is not possible to charge or change the battery of nodes. Thus energy efficiency and stability of network are the major research issues in wireless sensor network. Cluster based routing algorithms are good for increasing the energy efficiency, stability and performance of the sensor network.

In cluster based scheme sensor network is divided in to a number of clusters and a set of nodes from these clusters are periodically elected as the head of the cluster (CH) and these cluster heads collects the data from non-CHs nodes, aggregates the data and then send it to the base station [2] [3].Thus clustering is an important concept for evenly distributing the energy load in the sensor network. In heterogeneous network some of the nodes have more energy than the other nodes, which is the source of heterogeneity. DBCP is a novel energy efficient distance based cluster protocol for heterogeneous network. DBCP suggests a new Cluster head (CH) election mechanism based on the initial energy of nodes and the average distance of the nodes from the sink. The protocol is an improved version of SEP protocol presented in [7] and simulation result shows that it is more efficient than SEP.

The rest of this paper is organized as follows. Section 2 presents the related works. Section 3 describes the heterogeneous model and radio energy model, Section 4 describes DBCP protocol, Section 5 explores the simulation results and finally paper is concluded in Section 6.

## 2. RELEATED WORK

Clustering schemes are of two types: Homogeneous and Heterogeneous. In homogeneous clustering all the nodes have same amount of energy and in heterogeneous schemes nodes have different amount of energy. Designing cluster based routing schemes for heterogeneous network is a difficult task [9]. As a result most of the clustering schemes proposed for the wireless sensor network are homogeneous such as LEACH [4], HEED [6], and PEGASIS [8].

In clustering schemes cluster head have to spend more energy as they have double responsibility to get the data from the non cluster nodes, aggregates the data and then send it to the sink. For example LEACH [4] periodically assigns the duty of cluster head to different nodes and distributes the energy load by rotating this duty. LEACH [4] is good for homogeneous network but its performance is degraded in heterogeneous network [7].

HEED [6] is another important distributed clustering algorithm for homogeneous network in which residual energy of the nodes is used as main criteria for electing the cluster head. When HEED is applied to the heterogeneous network low energy sensor nodes may have higher election probability than the higher energy nodes. In PEGASIS [8] nodes are organized to form a chain and this chain can be either computed by the sensor nodes or by sink. Global knowledge of the network topology makes the implementation of PEGASIS difficult.

In [7] authors have proposed a new heterogeneous aware sensor network protocol named SEP. SEP elects cluster head based on the initial energy of nodes and it is a protocol for two level heterogeneous network. DEEC [9] proposed an algorithm in which cluster head is selected on the basis of probability ratio of residual energy of the node and average energy of the network.

In [10] strengths and weakness of many existing and new protocols are analyzed. EEPSC [10] divides the network into many static clusters and use cluster heads to distribute the load among high energy nodes. EEHCA [11] uses a backup concept for cluster heads to increase network life. EDGA [12] uses weighted election probabilities to elect cluster head for handling the heterogeneity of the network. EECDA) [13] uses a heterogeneous aware network model and maximum residual energy path for data transmission. In [14] authors have proposed a new distance based scheme for cluster head selection for increasing the network life.

## 3. HETEROGENEOUS NETWORK AND RADIO ENERGY DISSIPATION MODEL

### 3.1 Heterogeneous Network Model

This section describes the heterogeneous wireless sensor network model used in the paper. Network model consists of N sensors which are randomly deployed in a 100 X 100 square meters region as shown in Figure 1. Some of the assumptions made about the network model and sensors are as follows:

- Base station is located in the middle of the sensor field.
- Base station and nodes are stationary after deployment.
- Nodes continuously sense the region and they always have the data to send to the base station.
- Nodes do not have any knowledge about their location i.e. they are location unaware.
- Some percentage of the nodes have high energy then the other nodes
- Due to the harsh environment condition it is not possible to recharge the batteries of the nodes.

As described in [7] energy dissipated by the cluster head in a round is given by the Equation (1)

$$E_{CH} = (\frac{n}{k}-1)LE_{ele} + \frac{n}{k}LE_{DA} + LE_{elec} + L\varepsilon_{fs}d_{toBS}^2 \quad (1)$$

Where k is the number of clusters, $E_{DA}$ is the data aggregation and $d_{toBS}$ is the average distance between the cluster head and the sink. The energy dissipated by a non-CH node is given by the Equation (2)

$$E_{NCH} = L*(E_{elec} + \varepsilon_{fs}*d^2_{CH}) \quad (2)$$

The total energy dissipated in a cluster per round is given by Equation (3).

$$E_{Total} = E_{CH} + E_{NCH} \quad (3)$$

By substituting the Equation (1) and Equation (2) in Equation (3) we can find out the energy dissipated during a round which is given by the Equation (4)

$$E_{Total} = L(2nE_{elec} + nE_{DA} + \varepsilon_{fs}(kd_{toBS}^2 + nd_{CH}^2)) \quad (4)$$

Optimal no of clusters can be obtained by differentiating $E_{Total}$ with respect to k and putting it equal to zero.

$$k = \frac{\sqrt{\varepsilon_{fs}}}{\sqrt{\varepsilon_{mp}}} \frac{\sqrt{n}}{\sqrt{2\pi}} \frac{M}{d^2_{BS}} \quad (5)$$

According to [7] the average distance from cluster head to the sink is given by Equation (6)

$$d_{toBS} = 0.765 \frac{M}{2} \quad (6)$$

The optimal probability ($p_{opt}$) of a node to become cluster head is given by the Equation (7)

$$p_{opt} = \frac{1}{0.765}\sqrt{\frac{n}{2\pi}}\sqrt{\frac{\varepsilon_{fs}}{\varepsilon_{mp}}} \quad (7)$$

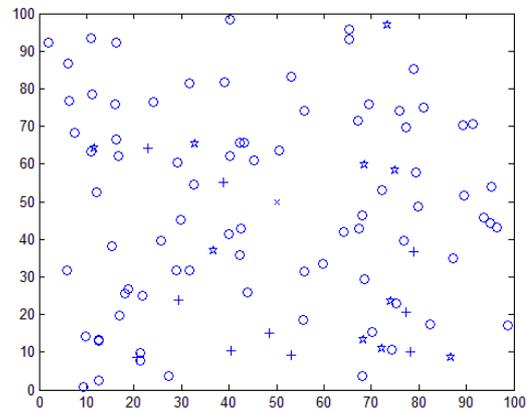

**Figure.1 100 nodes randomly deployed in the network(o normal node ,+ advanced node, * Super Nodes, x BS)**

### 3.2 Radio Energy Dissipation Model

This paper use the Radio energy model as described in [4]. Both free space ($d^2$ power loss) and the multipath fading ($d^4$ power loss) channel model is used depending upon the distance between the transmitter and receiver. If the distance is less than a particular threshold value then free space model are used otherwise multipath loss model is used. The amount of energy required to transmit L bit packet over a distance, d is given by Equation (8).

$$E_{TX(L,d)} = \begin{array}{l} L*E_{elec} + L*\varepsilon_{fs}*d^2 \quad if(d < d_o) \\ L*E_{elec} + L*\varepsilon_{mp}*d^4 \quad if(d \geq d_o) \end{array} \quad (8)$$

$E_{elec}$ is the electricity dissipated to run the transmitter or receiver circuitry. The parameters $\varepsilon_{mp}$ and $\varepsilon_{fs}$ is the amount of energy dissipated per bit in the radio frequency amplifier according to the distance $d_0$ which is given by the Equation (9).

$$d_o = \sqrt{\frac{\varepsilon_{fs}}{\varepsilon_{mp}}} \quad (9)$$

For receiving an L bit message the radio expends the energy given by Equation (10)

$$E_{RX}(L) = L * E_{elec} \quad (10)$$

## 4. DBCP PROTOCOL

DBCP contains three types of sensor nodes (i.e. normal, advanced and super) which are randomly deployed in the sensing region. Let m be the fraction of advanced nodes among normal nodes and $m_0$ is the fraction of super nodes among advanced node [13]. Let us assume that initial energy of the normal node is $E_0$. The initial energy each of the advanced and super nodes are $E_0(1+a)$ and $E_0(1+b)$ where a, b means that advanced and super nodes have **a** and **b** times more energy than the normal nodes. The total initial energy of the heterogeneous network is given by:

$$E_{Total} = N \cdot (1-m) \cdot E_o + N \cdot (m-m_o) \cdot E_o(1+a) + N m_o \cdot E_o(1+b)$$

$$= N \cdot E_o (1 + a \cdot (m - m_0) + m_0 \cdot b) \quad (11)$$

Thus the system has $(1 + a \cdot (m - m_0) + m_0 \cdot b)$ times more energy due to heterogeneous nodes. Normal, advanced and super nodes weighted probabilities are given by Equations (12-15).

$$p_n = \frac{p_{opt}}{1 + a.(m - m_0) + b.m_0} \quad (12)$$

$$p_a = \frac{p_{opt}}{1 + a.(m - m_0) + b.m_0} \times (1 + a) \quad (13)$$

$$p_s = \frac{p_{opt}}{1 + a.(m - m_0) + b.m_0} \times (1 + b) \quad (14)$$

As per [13] the threshold for the normal, advanced and super nodes is given by Equation (15),(16),(17)

$$T(n) = \begin{cases} \dfrac{p_n}{1 - p_n \times r(\text{mod } \frac{1}{p_n})} & \text{if } S_n \, \varepsilon \, G' \\ 0 & \text{otherwise} \end{cases} \quad (15)$$

$$T(a) = \begin{cases} \dfrac{p_a}{1 - p_a \times r(\text{mod } \frac{1}{p_a})} & \text{if } S_a \, \varepsilon \, G'' \\ 0 & \text{otherwise} \end{cases} \quad (16)$$

$$T(s) = \begin{cases} \dfrac{p_s}{1 - p_s \times r(\text{mod } \frac{1}{p_s})} & \text{if } S_s \, \varepsilon \, G''' \\ 0 & \text{otherwise} \end{cases} \quad (17)$$

The minimum required amplifier energy is proportional to the square of the distance from the transmitter to the receiver. Thus consumption of transmission energy is increased as the transmission distance increases. Cluster heads which are away from the base station require more energy to send the data to the BS. Thus far cluster heads require more energy to send data to BS than the CHs which are nearer to BS.

As a result there is a considerably difference between the energy consumption of the nodes which are nearer to the base station and then those which are far from the base station [14]. To remove this problem DBCP proposes a new distance based probability scheme so that the nodes which are far from the base station have the low chance to become cluster head. Initially after deployment base station broadcasts a signal to all the sensor nodes at a certain power level. Based on this signal strength each node computes its approximate distance from the BS.

Let $D_i$ is the distance between node $S_i$ and the base station. The average distance $D_{avg}$ [14] of the nodes can be calculated by using the Equation (18)

$$D_{avg} = \frac{1}{n} \sum_i^n D_i \quad (18)$$

According to [14] the Value of $D_{avg}$ can be approximated as

$$D_{avg} \simeq d_{TOCH} + d_{TOBS} \quad (19)$$

where $d_{TOCH}$ is the average distance between the node and the associate cluster head. $d_{TOBS}$ is the average distance between the cluster head and the sink.

Now if the distance $D_i < D_{avg}$ then DBCP use the following equation for calculating the threshold value T(n),T(a),T(s)

$$T(n) = \begin{cases} \dfrac{p_n}{1 - p_n \times (r \bmod \frac{1}{p_n})} \times (1 - \dfrac{D_i}{D_{avg}}) & \text{if } S_n \, \varepsilon \, G' \\ 0 & \text{otherwise} \end{cases} \quad (20)$$

$$T(a) = \begin{cases} \dfrac{p_a}{1 - p_a \times (r \bmod \frac{1}{pa})} \times (1 - \dfrac{D_i}{D_{avg}}) & \text{if } S_a \, \varepsilon \, G'' \\ 0 & \text{otherwise} \end{cases} \quad (21)$$

$$T(s) = \begin{cases} \dfrac{p_s}{1 - p_s \times (r \bmod \frac{1}{p_s})} \times (1 - \dfrac{D_i}{D_{avg}}) & \text{if } S_s \, \varepsilon \, G''' \\ 0 & \text{otherwise} \end{cases} \quad (22)$$

And if the distance $D_i >= D_{avg}$

Then the system will use the Equation (15), (16), (17) for finding the threshold value of the normal, advanced and super nodes.

## 5. SIMULATION RESULS

We have compared the performance of DBCP with LEACH and SEP. For evaluation we have used 100 x 100 square meters region with 100 sensor nodes as shown in Fig.1.Base station is located in the middle of the sensor field. We denote the normal nodes by using the symbol (o), advanced notes with (+), super nodes by (*) and the Base Station by (x). Radio parameters used for the simulations are given in Table 1. The performance metrics used for evaluating the protocols are:

(i) Network Lifetime: this is the time interval from the start of the operation till the last node alive

(ii) Stability Period: this is the time interval from the start of the operation until the death of the first alive node

(iii) Number of Alive Nodes per round

(iv) Throughput: No of packets send to Base station

DBCP is tested by introducing various parameters of heterogeneity. Figure 2 shows that network lifetime of DBCP is more than LEACH and SEP as first and last node dies later in DBCP as compared to LEACH and SEP. Figure 3 shows that no of alive nodes are more in DBCP as compared to LEACH and SEP. Figure 4 shows that number of packets send to Base station is more in DBCP as compared to LEACH and SEP. Thus throughput of DBCP is more than SEP and LEACH.

**Table 1 Radio Parameters used in Model**

| Parameter | Value |
| --- | --- |
| $E_{elec}$ | 5 nJ/bit |
| $\varepsilon_{fs}$ | 10 pJ/bit/m$^2$ |
| $\varepsilon_{mp}$ | 0.0013 pJ/bit/m$^4$ |
| $E_0$ | 0.5 J |
| $E_{DA}$ | 5 nJ/bit/message |
| Message Size | 4000 bits |
| $P_{opt}$ | 0.1 |
| do | 70m |

## 6. CONCLUSIONS

DBCP is a protocol for three level heterogeneous networks for taking the full advantage of heterogeneity. It improves the network lifetime, stable region and throughput of the network. To increase the energy efficiency of the network DBCP introduces a new distance based probability scheme in the system so that nodes which are near to the base station have the higher chances to become the cluster head. In this paper most of the aspects of the cluster based heterogeneous network have considered however future work includes introducing some level of mobility in the network

.

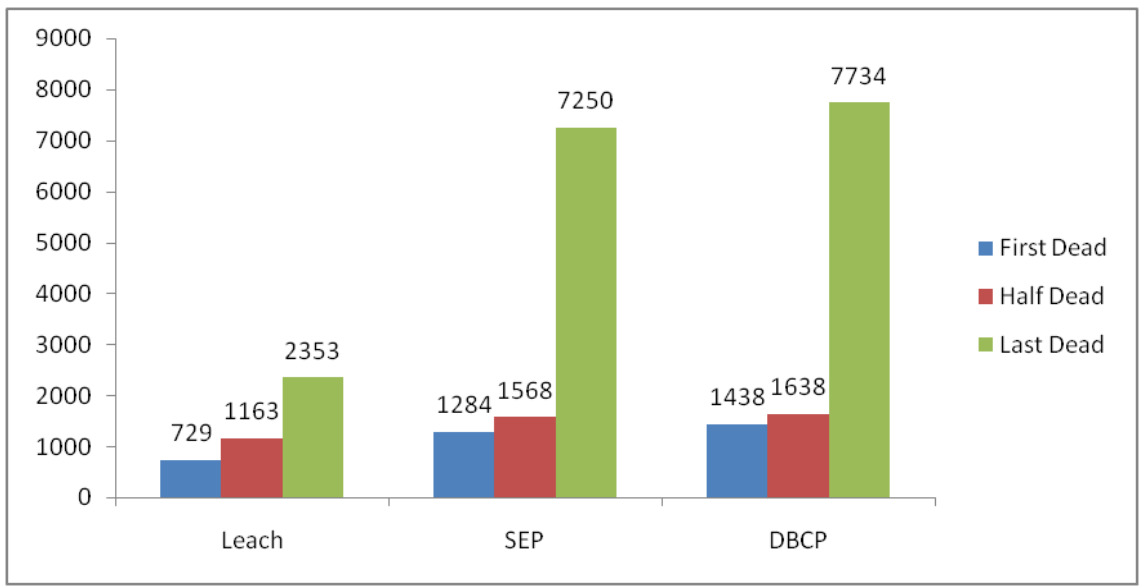

**Figure 2 (First, Half and Last Node Dead)**

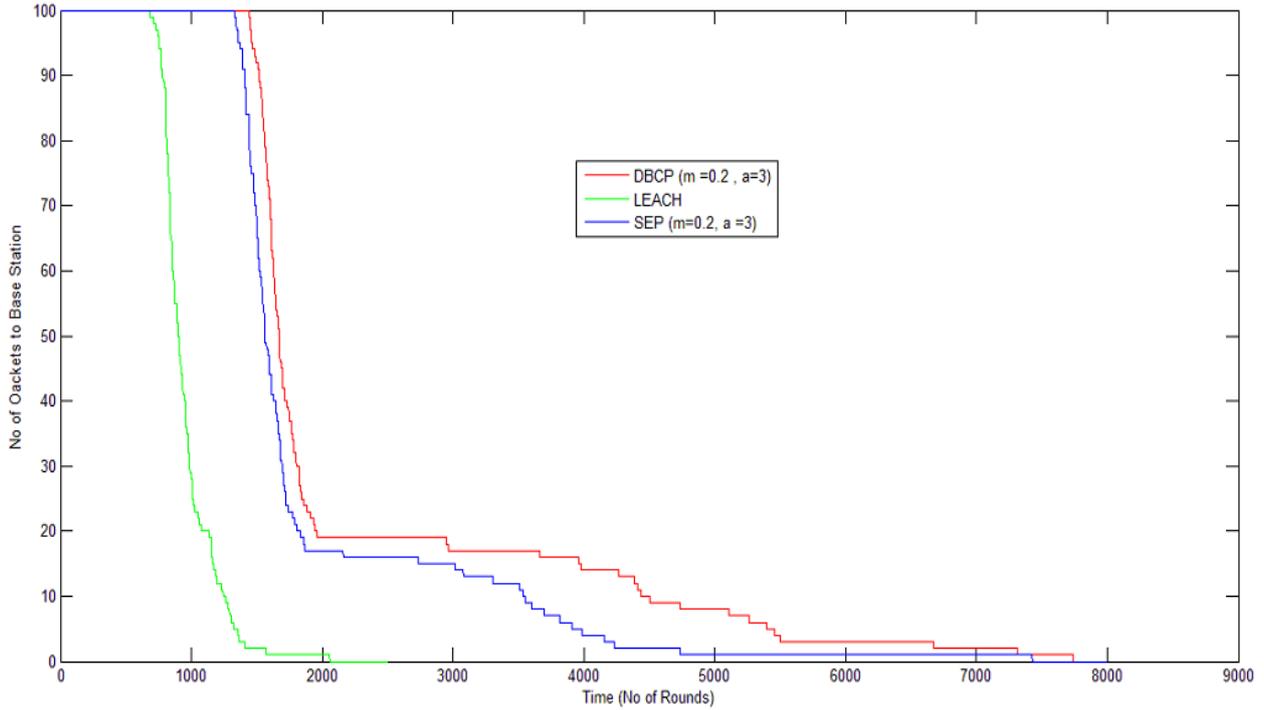

**Figure 3 (No. of Alive Nodes per round)**

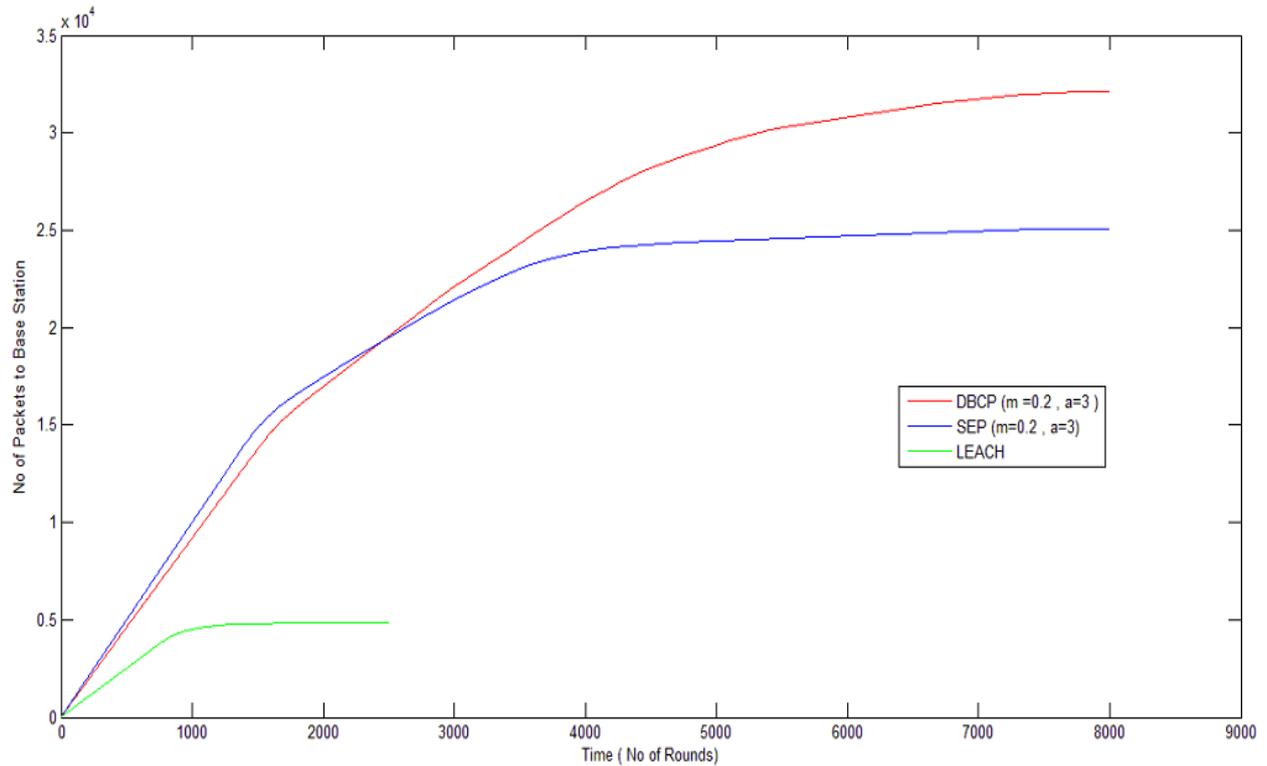

**Figure 4 (No. of Packets send to Base Station)**

## 7. REFERENCES

[1] I. F. Akyildiz, W.Su, Y. Sankarasubramaniam, and E. Cayirci, "A survey on sensor networks", IEEE Communications Magazine, vol. 40, no. 8 pp. 102–114, Aug. 2002

[2] Kemal Akkaya, Mohamed Younis, "A Survey on routing protocols for Wireless Sensor Networks", Ad Hoc Networks 3 (2005) 325-349, 2005

[3] S.K. Singh, M.P. Singh, and D.K. Singh, "A survey of Energy-Efficient Hierarchical Clustered Routing in Wireless Sensor Networks", IJANA, Sept.–Oct. 2010, vol. 02, issue 02, pp. 570–580 , 2010

[4] W. Heinzelman, A. Chandrakasan, H. Balakrishnan, "An application specific protocol architecture for Energy-efficient for wireless sensor networks", IEEE Transactions on Wireless Communications, 1(4), 660-670, 2002

[5] S. Bandyopadhyay, E.J. Coyle, "An Energy Efficient Hierarchical Clustering Algorithm for Wireless Sensor Networks," in: Proceeding of INFOCOM 2003, April 2003

[6] O. Younis, S. Fahmy, "HEED: A Hybrid, Energy-Efficient, Distributed clustering approach for Ad Hoc sensor networks", IEEE Transactions on Mobile Computing 3 (4) (2004), 366–379, 2004

[7] G. Smaragdakis, I. Matta, A. Bestavros, "SEP: A Stable Election Protocol for clustered heterogeneous wireless sensor networks", in: 2nd International Workshop on Sensor and Actor Network Protocols and Applications (SANPA 2004), 2004.

[8] S. Lindsey, C. Raghavendra,"PEGASIS: Power-Efficient Gathering in Sensor Information Systems," IEEE Aerospace Conference Proceedings, 2002, Vol. 3. No. 9-16, pp. 1125-1130, 2002

[9] L. Qing, Q. Zhu, M. Wang, "Design of a distributed energy-efficient clustering algorithm for heterogeneous wireless sensor networks". ELSEVIER, Computer Communications 29, pp 2230-2237, 2006

[10] S. Xun, "A Combinatorial Algorithmic Approach to Energy Efficient Information Collection in Wireless Sensor Networks", ACM Transactions on Sensor Networks, 3(1), 2007

[11] G. Xin, W.H. Yang, D. DeGang, "EEHCA: An Energy-Efficient clustering Algorithm for Wireless Sensor Networks", Information Technology Journal, 7(2):245-252, 2008

[12] Y. Mao, Z. Liu, L. Zhang, X. Li, "An Effective Data Gathering Scheme in Heterogeneous Energy Wireless Sensor Networks", Proceedings of International Conference on Computational Science and Engineering, 338-343, 2009

[13] Dilip Kumar, Trilok C. Aseri, R.B. Patel, "EECDA: Energy efficient clustering and data aggregation Protocol for Heterogeneous Wireless Sensor Network. Heterogeneous", IJCC & Control, Vol 6 (2011) No 1 pp 113-124, 2011

[14] Said Benkirane, Abderrahim Benihssane, M.Lahcen Hasnaoui, Mohamed Laghdir. "Distance-based Stable Election Protocol (DB-SEP) for Heterogeneous Wireless Sensor Network". IJCA, Nov. 2012